\documentclass[conference]{IEEEtran}
\pdfoutput=1
\IEEEoverridecommandlockouts
\usepackage{cite}
\usepackage{amsmath,amssymb,amsfonts}
\usepackage{algorithmic}
\usepackage{graphicx}
\usepackage{textcomp}
\usepackage{xcolor}
\usepackage{url}
\usepackage{hyperref}
\usepackage{mathabx, bm}
\usepackage{enumitem}
\usepackage{booktabs}
\usepackage{subfigure}
\usepackage{multicol,multirow}

\allowdisplaybreaks

\def\BibTeX{{\rm B\kern-.05em{\sc i\kern-.025em b}\kern-.08em
    T\kern-.1667em\lower.7ex\hbox{E}\kern-.125emX}}
\begin{document}

\title{QuantumCircuitOpt: An Open-source Framework for Provably Optimal Quantum Circuit Design\\
}

\author{\IEEEauthorblockN{Harsha Nagarajan}
\IEEEauthorblockA{\textit{Applied Mathematics and Plasma Physics (T-5)} \\
\textit{Los Alamos National Laboratory}\\
Los Alamos, NM, USA \\
harsha@lanl.gov}
\and
\IEEEauthorblockN{Owen Lockwood}
\IEEEauthorblockA{\textit{Department of Computer Science} \\
\textit{Rensselaer Polytechnic Institute}\\
Troy, NY, USA \\ 
lockwo@rpi.edu}
\and
\IEEEauthorblockN{Carleton Coffrin}
\IEEEauthorblockA{\textit{Advanced Network Science Initiative} \\
\textit{Los Alamos National Laboratory}\\
Los Alamos, NM, USA \\ 
cjc@lanl.gov}
}

\maketitle

\begin{abstract}
In recent years, the quantum computing community has seen an explosion of novel methods to implement non-trivial quantum computations on near-term hardware. 
An important direction of research has been to decompose an arbitrary entangled state, represented as a unitary, into a \textit{quantum circuit}, that is, a sequence of gates supported by a quantum processor.
It has been well known that circuits with longer decompositions and more entangling multi-qubit gates are error-prone for the current noisy, intermediate-scale quantum devices. To this end, there has been a significant interest to develop heuristic-based methods to discover compact circuits. 
We contribute to this effort by proposing \textit{QuantumCircuitOpt (QCOpt)}, a novel open-source framework which implements mathematical optimization formulations and algorithms for decomposing arbitrary unitary gates into a sequence of hardware-native gates.
A core innovation of QCOpt is that it provides \textit{optimality guarantees} on the quantum circuits that it produces.
In particular, we show that QCOpt can find up to 57\% reduction in the number of necessary gates on circuits with up to four qubits, and in run times less than a few minutes on commodity computing hardware. We also validate the efficacy of QCOpt as a tool for quantum circuit design in comparison with a naive brute-force enumeration algorithm.
We also show how the QCOpt package can be adapted to various built-in types of native gate sets, based on different hardware platforms like those produced by IBM, Rigetti and Google.
We hope this package will facilitate further algorithmic exploration for quantum processor designers, as well as quantum physicists.
\end{abstract}

\section{Introduction}
\label{sec:intro}
In the last decade, quantum computing has progressed from a research experiment to a tool on the brink of transforming a variety of computing-intensive industries, including medicine and transportation, to name a few \cite{preskill2018quantum}. Quantum computers have achieved rapid DNA sequencing and precise future traffic volumes prediction in urban areas \cite{kelly2018preview}. More importantly, quantum computing has also been recognized to play significant role in the United States' economic and defense infrastructure applications as stated in the 2018 national strategic overview \cite{qis_gov_2018}. The quantum computing market is projected to reach \$65 billion by 2030 from just \$507 million in 2019.

Perhaps, the best known application of quantum computing is to factor integers, whereas the fastest known classical algorithm is super-polynomial, while Shor's algorithm solves the problem in polynomial time on a quantum computer \cite{shor1999polynomial}. Quantum computers were also proposed for simulating Hamiltonian dynamics and studying phenomena in condensed-matter and high-energy physics \cite{feynman2018simulating}. These applications have motivated significant efforts toward building scalable quantum algorithms, typically modeled in the formalism of quantum circuits. Quantum circuits describe a computation as an ordered sequence of elementary quantum logic gates, or hardware-native gates, acting on quantum data, such as qubits. It is well known that any $n$-qubit quantum computation, represented as a target gate, can be achieved using a sequence of one- and two-qubit quantum logic gates \cite{barenco1995elementary}.
There are many ways of implementing the target gate using available hardware-native gates, and it is advantageous to find an implementation with minimum number of gates to migrate the negative impacts of noise and decoherence in a quantum computer's computation.
Of particular interest is to minimize the two-qubit CNOT gates, as they are orders-of-magnitude more error-prone than implementations of single-qubit gates. However, given a set of native gates, even for two-qubits, finding such optimal implementations with theoretical guarantees can be forbiddingly complex \cite{vatan2004optimal}.
While heuristics based on machine learning have been shown to achieve good circuits empirically \cite{khatri2019quantum, nam2018automated, younis2020qfast}, they do not provide a mechanism to measure how close they are to the best possible circuit.
The community is lacking both methods and software packages that can provide theoretical guarantees on the solution quality of quantum circuit decompositions.
\textit{To close this gap, this work proposes the software package QuantumCircuitOpt\footnote{\url{https://github.com/harshangrjn/QuantumCircuitOpt.jl}}, the first software that computes quantum circuit decompositions with numerically provable optimality guarantees, to the best of our knowledge.}
The key to QuantumCircuitOpt's success is to formulate the quantum circuit design task as carefully crafted mixed-integer programs (MIPs) that can be solved with state-of-the-art discrete optimization software \cite{bixby2004mixed}.
Recognizing that different hardware vendors implement different hardware-native gates, QuantumCircuitOpt seamlessly adapts the design task to different user-specified gates. Through these features, QuantumCircuitOpt makes sophisticated mathematical optimization technologies accessible to all quantum computing users.

\section{QuantumCircuitOpt's Framework}
\label{sec:qcopt}

\subsection{Success of mixed-integer programming}
\begin{equation}
		\label{eq:mip}
		\begin{array}{ll}
	  		 \mathrm{minimize} &   c^T z, \hspace{1.78cm} (\mathrm{objective}) \\
		\mathrm{subject \ to} &  Ax + Cz = b, \quad (\mathrm{linear \ constraints}) \\
			& x_l \leq x \leq x_u, \quad \ (\mathrm{bound \ constraints}) \\ 
			& z \in \{0, 1\}^{|z|}. \quad \hspace{0.31cm} (\mathrm{integer \ constraints})
		\end{array}
\end{equation}
Mixed-integer programming (MIP), a special case falling under an umbrella of mathematical programming, has been an integral part of important decision making problems such as the classic traveling salesman problem, airline scheduling and energy systems optimization, to name a few \cite{wolsey2007mixed}. The mathematical theory of MIP, shown in \eqref{eq:mip}, is concerned with finding amongst all solutions of a system of linear algebraic equalities and inequalities, written as functions of decision variables, a variable assignment that minimizes a linear function of those variables. The variables in an MIP can be a mix of  continuous and integer values. Model \eqref{eq:mip} represents a canonical form of an MIP, where $x$ and $z$ are the continuous and integer variables, respectively. Solving MIPs, while NP-Hard in the worst case, has made dramatic strides in practical applications via state-of-the-art commercial solvers such as CPLEX and Gurobi, where decades of research and development (R\&D) have yielded algorithmic performance improvements that even outpaced the hardware improvements from Moore's Law. For example, CPLEX, since its release in 1991, has seen $\approx$200 billion factor speedup in solving MIPs \cite{bertsimas2014statistics}. 

At their core, MIPs are declarative, they specify what optimization problem to solve and not how to solve it. Hence, a performant mathematical programming language is required to access MIP solvers and to enable a separation from the problem-specific complexities and the underlying algorithms employed by such solvers.
For this purpose, JuMP is one of the most widely adopted packages for mathematical programming and provides a declarative domain specific language (DSL) in the Julia programming language for modeling mathematical programs \cite{dunning2017jump}, such as MIPs. 
By taking a ``leap of faith'' that the problem of optimal quantum circuit design can also be posed as an MIP like shown in \eqref{eq:mip} (details in section \ref{sec:math_models}), we now present the framework of the QuantumCircuitOpt software package to formulate and solve such problems. 

\subsection{Features of the QCOpt package}
\label{subsec:qcopt_features}
\begin{figure}
    \centering
    \includegraphics[scale=0.17]{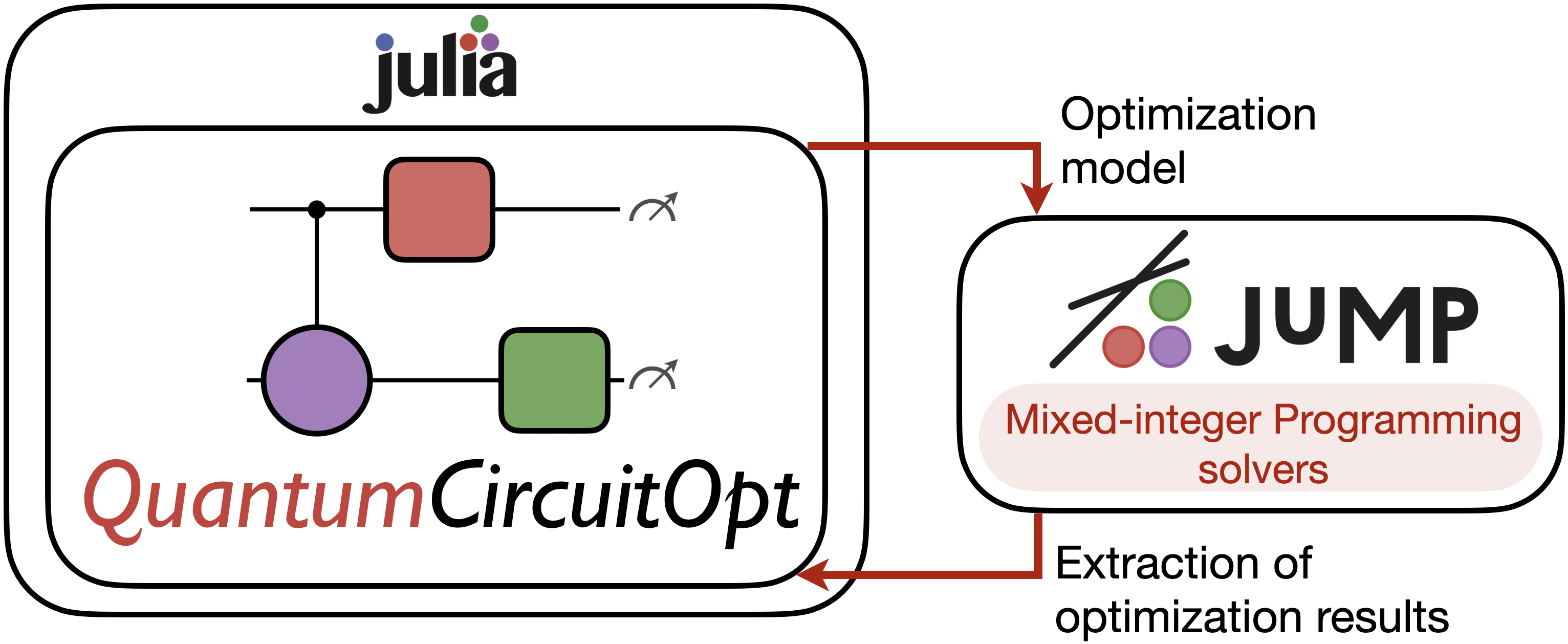}
    \caption{Illustration of QCOpt's framework for converting QuantumCircuitModel into a JuMP model.}
    \label{fig:framework}
\end{figure}

Building on the recent success of Julia, JuMP and mixed-integer programming, in this work we introduce \textit{QuantumCircuitOpt} (or QCOpt), a free and open-source toolkit for quantum circuit design. As illustrated in Figure \ref{fig:framework}, QCOpt is written in Julia, a relatively new and fast dynamic programming language used for technical computing with support for extensible type system and meta-programming \cite{bezanson2017julia}. At a high level, QCOpt provides an abstraction layer to achieve two primary goals: (a) to capture user-specified inputs, such as a desired quantum computation and the available hardware gates, and build a JuMP model of an MIP formulation and (b) to extract, analyze and post-process the solution from the JuMP model to provide exact and approximate circuit decompositions, up to a global phase and machine precision. In the remainder of this section, we will highlight a few important features of QCOpt in detail. 
\\ \\ \textbf{User inputs} \\ 
QCOpt's user inputs is a dictionary, with a collection of key-value pairs, where the keys are the options to be set and values are the user inputs. These key-value pairs need not be of the same data type, as seen below. While the below list is not comprehensive, it provides sufficient information to start exploring the package: 
\begin{enumerate}[leftmargin=*]
    \item \texttt{"num\_qubits"}: Total number of qubits of the circuit.  
    
    \item \texttt{"maximum\_depth"}: Maximum allowable depth for the circuit's decomposition ($\geq$ 2). From QCOpt's perspective, maximum depth is also the total number of input \texttt{elementary\_gates} (see below) allowed in the decomposition. 
    
    \item \texttt{"elementary\_gates"}: A vector of all one- and two-qubit elementary gates, native to the hardware. For example, on a three qubits circuit, here is an allowable input: \texttt{["H\_1", "H\_2", "T\_2", "Tdagger\_3", "CNot\_1\_2", "CNot\_3\_2", "CNot\_3\_1", "Identity"]}. Notice that an one-qubit gate is associated with it's qubit location and a two-qubit gate is associated with it's control and target qubit locations. For example, inputs \texttt{"CNot\_1\_2"} and \texttt{"CNot\_3\_2"} implement the regular $\mathrm{CNOT}$ and the reverse $\mathrm{CNOT}$ gates in corresponding qubit locations. 
    
    \item \texttt{"target\_gate"}: A target unitary quantum gate which needs to be compiled/decomposed using the above-mentioned native \texttt{"elementary\_gates"}.
    
    \item \texttt{"objective"}: The options are (a) \texttt{"minimize\_depth"} to minimize the total number of one- and two-qubit gates in the decomposition, and (b) \texttt{"minimize\_cnot"} to minimize the total number of $\mathrm{CNOT}$ gates in the decomposition. 
    
    \item \texttt{"decomposition\_type"}: The options are: (a) \texttt{"exact"}: finds an exact, optimal circuit decomposition up to a global phase and machine precision, provided it exists, (b) \texttt{"approximate"}: Finds an approximate decomposition with best gate fidelity found within the limits of run time. 
    
    \item \texttt{"set\_cnot\_lower\_bound"}: While this is an optional input, it aids in reducing the size of the search space of MIP and also helps in exploiting theoretical bounds on $\rm CNOT$ gate costs from the literature \cite{shende2008cnot}.  
\end{enumerate}

\begin{figure}
    \centering
    \includegraphics[scale=0.77]{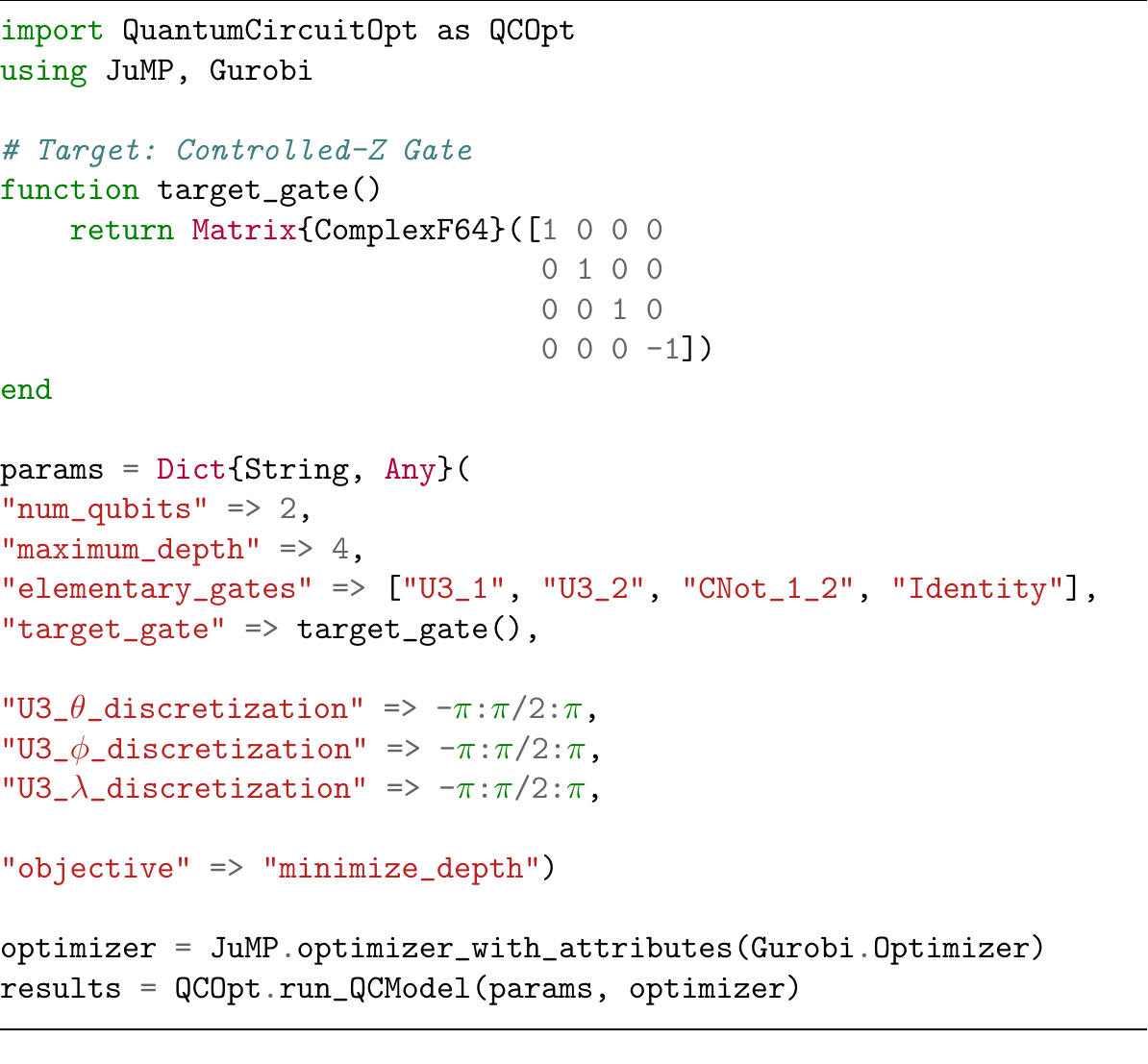}
    \vspace{-0.9cm}
    \caption{A sample implementation to decompose the controlled-$\rm Z$ gate using $\rm U_3$ and $\rm CNOT$ gates in QCOpt.}
    \label{code:czgate}
\end{figure}

Note that if a gate with continuous angles is in the input \texttt{elementary\_gates}, then discretization of these angles have to be specified in the inputs. For example, see figure \ref{code:czgate} for specifying the angles for the $\rm U_3$ gate.

Also, it is important to keep in mind that the optimization tasks handled by QCOpt are NP-hard to compute (in worst-case) in the size of the following set of inputs: \texttt{num\_qubits}, \texttt{maximum\_depth} and \texttt{elementary\_gates}. \\

\noindent
\textbf{Available gates set}

Currently, QCOpt supports the gates of the Clifford group ($\rm H, S, S^{\dagger}, X, \sqrt{X}, \sqrt{X}^{\dagger}, Y, Z, CNOT$), and the $\pi/8$ phase gate, $\rm T$, and it's conjugate transpose, $\rm T^{\dagger}$, which together are sufficient to arbitrarily approximate 
any target gate with a finite set of operations. However, for enhanced generality, users can also input universal single-qubit operators such as the rotation gates, $\rm (R_x(\theta), R_y(\theta), R_z(\theta)$) and the universal gate with three Euler angles, $\rm U_3(\theta, \phi, \lambda)$, along with the corresponding angle discretizations. QCOpt also supports various two-qubit gates such as the $\rm Swap$, $\rm iSwap$, $\rm Magic$-basis, $\rm Grover$'s diffusion operator, $\rm Sycamore$, $\rm QFT2$ and controlled versions of all the single-qubit gates in the Clifford group,  $\rm T, T^{\dagger}, V, V^{\dagger}, R_x, R_y, R_z$ and $\rm U_3$ gates. QCOpt also supports Kronecker products of one- and two-qubit gates as a native gate in the following form: for e.g., on three-qubit circuits, ($\rm H_1 \otimes T_2 \otimes T_3$), ($\rm T_1 \otimes CNOT_{2,3}$). 

\noindent
\\ \textbf{Sample implementation}

Figure \ref{code:czgate} shows a sample implementation for decomposing a two-qubit controlled-$\rm Z$ gate. The native gates set contains discretized $\rm U_3$ gates which can be located on both the qubits, $\rm CNOT$ gate and the $\rm I$ gate. The maximum allowable depth for the decomposition is four and Gurobi is specified as an MIP solver. Finally, the results from the optimization model, including an optimal decomposition, run time and solver status, can be accessed from the \texttt{results} dictionary. QCOpt solves this optimization model to optimality, with 72 native gates (after discretization) per depth, in less than 4 seconds on a commodity computing hardware.

\subsection{Pre-Solving in QCOpt}
\label{subsec:preopt_enhance}
Pre-processing and transformation of mathematical programs before solving, known as pre-solving, is a well established procedure for improving run time of mathematical optimization algorithms \cite{achterberg2020presolve}.
To that end, before solving the main optimization model (MIP) to optimally, QCOpt preprocesses the user-defined inputs to make them more amenable to the MIP solver's algorithm. Here we present a few such pre-solving  techniques:  
\begin{enumerate}[leftmargin=*]
    \item Elimination of identical gates in the \texttt{"elementary\_gates"} input, including the gates obtained via discretization of angle parameters. 
    \item Implementation a compact, smaller-sized MIP formulation if both the input \texttt{"elementary\_gates"} and \texttt{"target\_gate"} contain purely real-valued elements.
    \item An optional user-given \texttt{"input\_circuit"} is preprocessed to map this readily available circuit to warm-start the MIP solver with a feasible decomposition, as this can improve solver run times. During this pre-processing step, if QCOpt recognizes the input circuit as a feasible solution, then it activates the option to emphasize on proving optimality than finding feasibility for the MIP solver. For example, in the Gurobi solver, the \texttt{MIPFocus} parameter is set to ``2'' to emphasize optimality. 
\end{enumerate}

\section{Underlying Mathematical Programs}
\label{sec:math_models}
\begin{figure}[ht]
    \centering
    \includegraphics[scale=0.17]{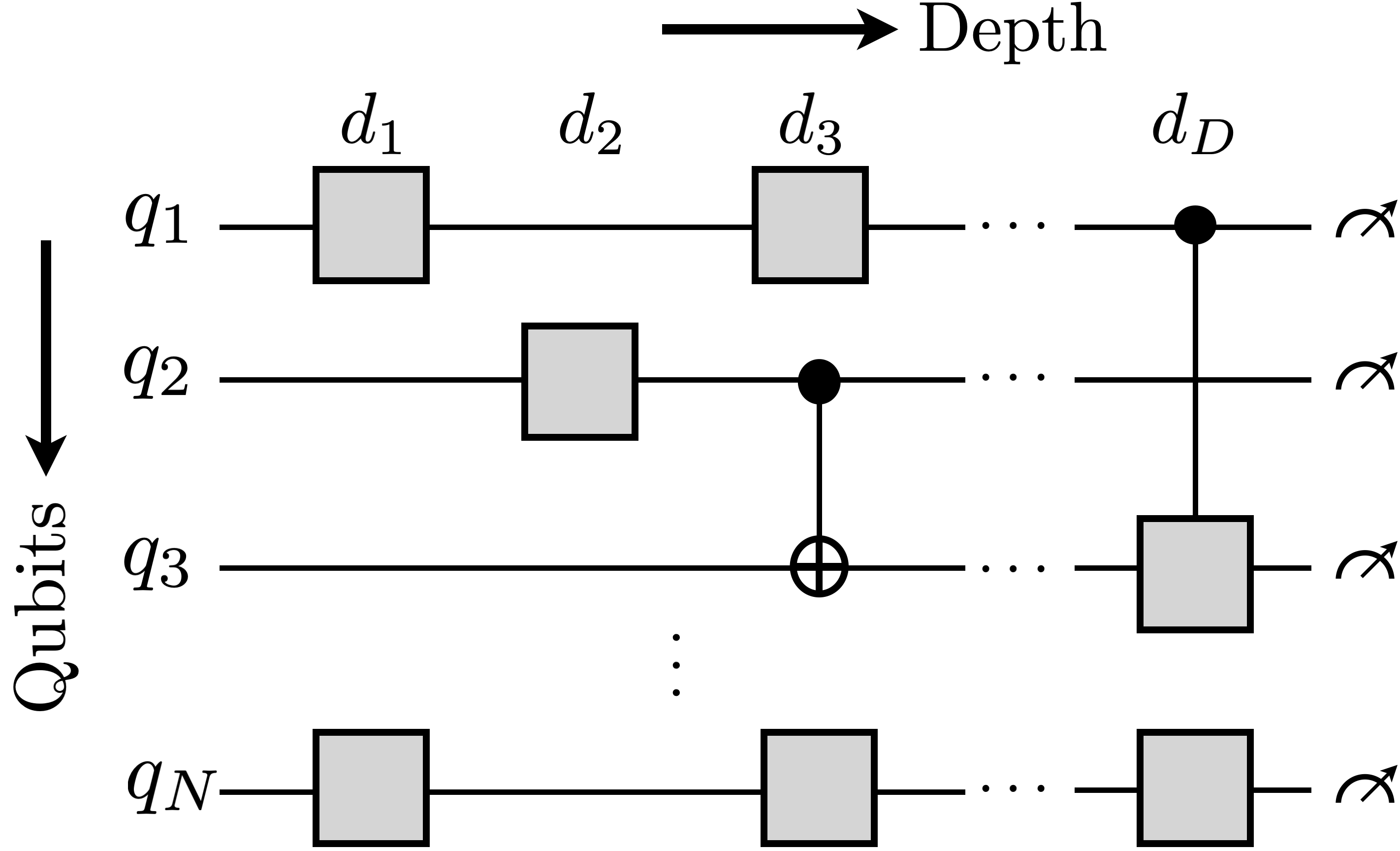}
    \caption{A generic quantum circuit of $N$ qubits and $D$ depth, with an ordered sequence of one- and two-qubit quantum logic gates.}
    \label{fig:circuit}
\end{figure}
QCOpt focuses on implementing compact and efficient MIP formulations for the design of a generic quantum circuit, represented as an ordered sequence of one- and two-qubit quantum gates as shown in Figure \ref{fig:circuit}. We now present a simplified derivation of the MIP formulation implemented in QCOpt. For clarity purposes, we consider IBM quantum experience's native gates set \cite{khatri2019quantum} in the following exposition. However, the formulation in QCOpt is general enough to support other native gates of a different quantum processor in a straight-forward fashion. 

Let the native gates set contain the following one- and two-qubit complex-valued gates:
\begin{subequations}
\begin{align}
\mathrm{U_3}(\hat{\theta}, \hat{\phi}, \hat{\lambda}) & =
    \begin{bmatrix}
        \cos(\hat{\theta})          & -e^{\mathrm{i} \hat{\lambda}}\sin(\hat{\theta}) \\
        e^{\mathrm{i}\hat{\phi}}\sin(\hat{\theta}) & e^{\mathrm{i}(\hat{\phi}+\hat{\lambda})}\cos(\hat{\theta})
    \end{bmatrix},  \\  
\mathrm{R_x}(\pi/2) &= \frac{1}{\sqrt{2}} 
\begin{bmatrix}
1 & -\mathrm{i}\\ 
-\mathrm{i} & 1
\end{bmatrix} \ 
\mathrm{I} = 
\begin{bmatrix}
1 & 0 \\
0 & 1
\end{bmatrix}, \\ 
\mathrm{CNOT}_{c,t} &= 
\begin{bmatrix}
1 & 0 & 0 & 0 \\
0 & 1 & 0 & 0 \\ 
0 & 0 & 0 & 1 \\
0 & 0 & 1 & 0
\end{bmatrix}, \label{eq:cnot_gate}
\end{align}
\label{eq:natives}
\end{subequations}
where $\mathrm{i}$ is the imaginary constant and $\hat{\theta}, \hat{\phi}$ and $\hat{\lambda}$ represent discretized angles of continuous angle parameters of the universal $\mathrm{U_3}$ gate. From the IBM quantum processor perspective, we consider a special case of this gate, i.e, the phase gate, with $\hat{\theta}$ and $\hat{\phi}$ from the set $\{0\}$ and $\hat{\lambda} \in \mathcal{L}$. For example, if $\mathcal{L} =  \{-\pi/2, -\pi/4, \pi/4, \pi/2\}$, then the total number of discretized $\mathrm{U_3}$ gates considered for optimization is four, in addition to the remaining three native gates with constant parameters. Notice that we discretize the angles instead of handling them as continuous variables in order to circumvent the resulting non-linear, non-convex terms in the optimization model. 
However, QCOpt's framework based on JuMP, makes it flexible to interact with existing Julia-based non-linear solvers \cite{kroger2018juniper}, which will be a future direction of work.  

In equation \eqref{eq:cnot_gate}, $c$ and $t$ represent control and target qubits, respectively. QCOpt follows the convention where higher qubit indices are more significant, and the lesser significant qubit serves as the control qubit for two-qubit controlled gates. 



Let $N$, $D$, and $\mathrm{T^g}$ be the number of qubits in the circuit, maximum allowable depth of the circuit and the target unitary gate to be decomposed by the optimization model, respectively. Next, we introduce binary variables to choose the location of the these native gates in the circuit: 
\begin{align*}
    & z^u_{n,d, k} \in \{0,1\}, \ z^r_{n,d} \in \{0,1\}, \ z^{cnot}_{n_c, n_t, d} \in \{0,1\}, \ z^{id}_{d} \in \{0,1\} \\ 
    & z^u_{n,d, k} = 1 \ \ \textit{iff} \ \ \mathrm{U_3}(\hat{\lambda}_k) \ \mathrm{is \ placed \ on \ qubit \ } n, \mathrm{depth} \ d  \\ 
    & z^r_{n,d} = 1 \ \ \textit{iff} \ \  \mathrm{R_x}(\pi/2) \ \mathrm{is \ placed \ on \ qubit \ } n, \mathrm{depth} \ d \\
    & z^{cnot}_{n_c,n_t,d} = 1 \ \ \textit{iff} \ \  \mathrm{CNOT}_{n_c, n_t} \ \mathrm{is \ placed \ on \ qubits \ } (n_c, n_t), \nonumber \\
    & \hspace{2cm} \mathrm{depth} \ d, \\ 
    & z^{id}_{d} = 1 \ \ \textit{iff} \ \  \mathrm{I} \ \mathrm{is \ placed \ at} \mathrm{ \ depth} \ d 
\end{align*}
where $d \in \{1,\ldots,D\}$, $n, n_c, n_t \in \{1,\ldots,N\}$, and an ordered pair $(n_c, n_t)$ represents control and target qubits on which the two-qubit $\mathrm{CNOT}_{n_c,n_t}$ gate can be placed. 

Next, let $\mathrm{G}_d$ represent the choice of one of the above-described native gates at every depth. Again, for simplicity of exposition, we present $\mathrm{G}_d$ in 2-qubits below. However, QCOpt implements the generalized version on $N$ qubits. For every depth $d$ in $1,\ldots, D$,
\begin{subequations}
\begin{align}
    \mathrm{G}_d =&  \sum_{\hat{\lambda}_k \in \mathcal{L}} z^{u}_{1,d,k} (\mathrm{U_3}(\hat{\lambda}_k) \otimes {\mathrm{I}}) +  \sum_{\hat{\lambda}_k \in \mathcal{L}} z^{u}_{2,d,k} ({\mathrm{I}} \otimes \mathrm{U_3}(\hat{\lambda}_k)) \nonumber \\ 
    & + z^{r}_{1,d} ( \mathrm{R_x}(\pi/2)  \otimes \mathrm{I}) 
      + z^r_{2,d} (\mathrm{I} \otimes \mathrm{R_x}(\pi/2)) \nonumber \\
    &  + z^{cnot}_{c_{1,2,d}} {(\mathrm{CNOT}_{1,2})} + z^{cnot}_{c_{2,1,d}} (\mathrm{CNOT}_{2,1}) + z^{id}_{d} ( \mathrm{I} \otimes \mathrm{I}), \\
    & \hspace{-0.8cm} \sum_{n=1}^{2} \left( \sum_{\hat{\lambda}_k \in \mathcal{L}} z^{u}_{n,d,k} + z^{r}_{n,d} \right) + z^{cnot}_{c_{1,2,d}} + z^{cnot}_{c_{2,1,d}} + z^{id}_{d}  = 1 \label{eq:Gd_cons_b}
\end{align}
\label{eq:Gd_cons}
\end{subequations}
where $\otimes$ represents the Kronecker product of one-qubit gates. Constraint \eqref{eq:Gd_cons_b} ensures that only one gate, out of all native gates, is placed per depth of the decomposition. Note that this constraint does not lose the generality of representing any quantum circuit. It is well known that the gates appearing in two consecutive depths and at non-identical qubits can be compressed into an equivalent single-depth circuit.  

The multiplicative property of unitary gates, ${G}_d$, on a quantum circuit which is intended to represent any computation, given by a target gate $\mathrm{T^g}$, can be mathematically modeled as the following constraint in an optimization model: 
\begin{align}
    \mathrm{G}_0 \prod_{d=1}^{D} \mathrm{G}_d = \mathrm{T^g}
    \label{eq:Gd_prod}
\end{align}
where $\mathrm{G}_0$ is an initial state of the circuit from which the computation evolves. 

The core contribution of QCOpt package also lies in an efficient implementation of the non-linear, non-convex constraint in \eqref{eq:Gd_prod}, as this can be crucial in the MIP solver's performance. Optimization tasks with such constraints with products of variable-matrices is known to be NP-hard to compute \cite{tran2017antibiotics}. Moreover, state-of-the-art MIP solvers, such as CPLEX and Gurobi, do not support constraints in the above-described form. Hence, we generalize the so-called ``recursive linearization'' technique to variable-matrix products, akin to handling a multilinear term of scalar variables in the context of deterministic global optimization \cite{nagarajan2019adaptive,nagarajan2016tightening,sundar2021piecewise}. Thus, QCOpt reformulates constraint \eqref{eq:Gd_prod} as follows: 
\begin{subequations}
\begin{align}
    & \widehat{\mathrm{G}}_d = \widehat{\mathrm{G}}_{d-1} \mathrm{G}_d \quad \forall d=2,\ldots,(D-1), \label{eq:recursive}\\ 
    & \widehat{\mathrm{G}}_1 = \mathrm{G}_0 \mathrm{G}_1, \quad \widehat{\mathrm{G}}_{d-1}\mathrm{G}_D = \mathrm{T^g}
\end{align}
\label{eq:recursive_all}
\end{subequations}
In the above reformulation, $\widehat{\mathrm{G}}_d$ represents cumulative products of variable unitary gates, thus preserving the property of being an unitary matrix. While the elements of  $\mathrm{G}_d$ matrix are affine functions of only binary variables (see constraints \eqref{eq:Gd_cons}), the elements of $\widehat{\mathrm{G}}_d$ matrix can be values in a complex field $\mathbb{C}$, whose real and imaginary parts lie within the box $[-1,1]$. Thus, the constraints in \eqref{eq:recursive} are still non-linear, with elements of $\widehat{\mathrm{G}}_d$ being bilinear products of a continuous and a binary variable. Let one such product term be $xz$ where $x \in [-1,1]$ represents either real or an imaginary part of an element of $\widehat{\mathrm{G}}_d$  and $z \in \{0,1\}$ represents a binary variable associated with any one of the native gates. This product can be efficiently and exactly linearized by applying the following set of four affine constraints with an auxiliary variable $\widecheck{xz}$ per product term: 
\begin{subequations}
\begin{align}
    \widecheck{xz} \geqslant -z, \quad &\widecheck{xz} \geqslant x+z-1, \\ 
    \widecheck{xz} \leqslant z, \quad &\widecheck{xz} \geqslant x-z+1.
\end{align}
\label{eq:mcc}
\end{subequations}
The above linearization reformulation of non-linear constraints in \eqref{eq:recursive_all}, also known as McCormick linearization \cite{mccormick1976computability}, is named the \texttt{compact\_formulation} in QCOpt. However, QCOpt also implements a theoretically stronger linearization formulation based on convex-hull disjunctions proposed by Balas from his theory of disjunctive formulations \cite{balas2018disjunctive}. Since a stronger formulation leads to a better linear programming relaxation of an MIP \cite{wolsey2007mixed}, this formulation was observed to be faster on smaller qubit circuits. Hence, QCOpt provides this alternative choice of the formulation, which is named the \texttt{balas\_formulation}.  

\subsection{Objective function}
\label{subsec:objective}
Currently QCOpt implements two separate objective functions  relevant in the field of quantum circuit design. However, the software can easily be extended to support other objectives as needed, such as minimizing the cross-talk noise on quantum processors \cite{murali2020software}. 
To be consistent with the previous derivations, we now present these functions on the two-qubit circuit: 

a) To minimize the total depth, which is also the total number of one- and two-qubit gates admitted in the decomposition, as this serves as a proxy for the execution time of the circuit. This objective is modeled as a linear function as follows:
\begin{align*}
    \mathrm{minimize} \ \sum_{d=1}^{D} \left(\sum_{n=1}^{2} \left( \sum_{\hat{\lambda}_k \in \mathcal{L}} z^{u}_{n,d,k} + z^{r}_{n,d} \right) + z^{cnot}_{c_{1,2,d}} + z^{cnot}_{c_{2,1,d}}\right)
    \label{eq:obj_1}
\end{align*}
b) To minimize the total number of $\mathrm{CNOT}$ gates in the decomposition, modeled as a linear function as follows: 
\begin{align*}
    \mathrm{minimize} \ \sum_{d=1}^{D} \left( z^{cnot}_{c_{1,2,d}} + z^{cnot}_{c_{2,1,d}} \right)
\end{align*}

Note that it is important to have an $\rm Identity$ gate ($\rm I$) in the native gates set in order to obtain a compact circuit decomposition while minimizing the above mentioned objectives. This ensures an optimal decomposition of a depth lesser than the given value of $D$, if such a decomposition is feasible to the optimization model. 

To summarize, the optimization model, with the linear objective function and the set of linearized constraints as described earlier in this section, is a mixed-integer program (MIP) which can be solved to global optimality using state-of-the-art MIP solvers. We now provide a brief discussion on the optimality guarantees of these solvers. 

\subsection{MIP optimality guarantees} 
\label{subsec:mip_opt_guarantees}
State-of-the-art solvers such as CPLEX and Gurobi are very well developed technologies with a solid mathematical footing and decades of R\&D towards solving generic MIPs with convex constraints. These solvers employ a combination of branch-and-bound (B\&B) algorithm and cutting plane generation to solve a MIP to global optimality \cite{lodi2010mixed,wolsey2007mixed}. In its basic form, a B\&B is a \textit{divide-and-conquer} approach that partitions an exponential search space sequentially into smaller sub-problems, also referred to as nodes of the B\&B tree, which are intended to be easier to solve. At each node of this B\&B tree, a linear programming (LP) relaxation is solved by relaxing the integrality of the binary variables. During this process of ``branching'', the LP relaxation's objective serves as a lower bound and an integral solution's objective, obtained using  heuristics, serves an upper bound to the optimal solution of the MIP.
If at any node, the LP relaxation value is greater than the best integral solution encountered so far, then the search can safely be stopped at that node as none of its children will yield a better solution than the current best solution.
Termination criterion for the B\&B algorithm is when the obtained LP relaxation is also integral, that is, when the lower and the upper bounds are close enough to each other, thus \textit{guaranteeing the global optimality of this integer solution}. Despite the theoretical complexity of this algorithm being exponential in the worst-case, it can be accelerated by orders-of-magnitude by incorporation of effective and valid constraints (cutting planes or cuts) into this scheme. State-of-the-art MIP solvers incorporate an array of such generic cuts, like Gomory, mixed-integer rounding and flow-cover cuts, to name a few \cite{wolsey2007mixed}. While these cuts perform well for generic MIPs, there still remain several technical insights and careful analysis which can lead to problem-specific cuts, which MIP solvers cannot generate automatically. Hence, in the forthcoming section, we summarize a few effective cuts which QCOpt implements to accelerate the convergence to global optimality.

\subsection{Search space reduction via valid inequalities}
\label{subsec:valid}
QCOpt implements various symmetry-breaking, valid constraints in form of inequalities which have been observed to be very critical in reducing the search space or the number of nodes explored in the B\&B tree. For example, these mathematical insights are shown in Section \ref{subsec:valid_cons_result} to provide up to a \textit{180 times improvement} in runtime of QCOpt even on two-qubit circuits. Here, we summarize these constraints: 
\begin{enumerate}[leftmargin=*]
    \item \textit{Commuting gates}: Given any two unitary native gates $\mathrm{U}_1$ and $\mathrm{U}_2$, they are said to commute if their commutator, $[\mathrm{U}_1, \mathrm{U}_2] = \mathrm{U}_1\cdot \mathrm{U}_2 - \mathrm{U}_2\cdot\mathrm{U}_1$ is equal to zero. Thus including the following inequality in a pairwise fashion will forbid the symmetric feasible solution $U_2U_1$, while not changing the optimal circuit: 
    $$z^{u_2}_d + z^{u_1}_{d+1} \leq 1 \quad \forall d=1,\ldots, D$$
    
    Commuting gate constraints can also be further generalized to deeper sub-circuits of equivalent patterns, which can be found extensively in the quantum literature \cite{nam2018automated,maslov2008quantum,dueck2018optimization}. For example, while the sequence ``$\mathrm{CNOT}_{1,2}\cdot \mathrm{H}_2 \cdot \mathrm{CNOT}_{2,3} \cdot \mathrm{H}_2$'' is a possible feasible sub-circuit, an equivalent sub-circuit ``$\mathrm{H}_2 \cdot \mathrm{CNOT}_{2,3} \cdot \mathrm{H}_2 \cdot \mathrm{CNOT}_{1,2}$'' can be eliminated from the search space via a valid inequality. QCOpt is currently equipped to identify such classic patterns and apply constraints to forbid equivalent sub-circuits. 
    
    \item \textit{Involutory gates}: Any unitary native gate, $\mathrm{U}$ is said to be involutory, if and only if $\mathrm{U}^2$ is an Identity ($\mathrm{I}$) gate. Assuming that $\mathrm{I}$ is already in the native gates set, such consecutive pairs of $\rm U$ gate can be eliminated using an inequality similar as above.   
    
    \item \textit{Idempotent gates}: Any unitary native gate, $\mathrm{U}$ is said to be idempotent, if and only if $\mathrm{U}^2 = \mathrm{U}$. Again, such consecutive pairs of $\rm U$ gate can be eliminated using an inequality. 
    
    \item \textit{Redundant gate-pairs}: Let the set of native gates be $\mathcal{U} = \{\rm U_1, U_2, \ldots, U_m\}$. Any gate-pair index which belongs to the set $\{(i,j) | (i\neq j) \wedge (\mathrm{U}_i\cdot \mathrm{U}_j \in \mathcal{U} \ \forall i,j = 1,\ldots,m) \}$ is considered as a redundant pair, and such pairs can be eliminated via  inequalities as described above. 

\end{enumerate}

\subsection{Proving Circuit Infeasibility in QCOpt}
\label{subsec:infeas}
In the process of compiling any arbitrary entangled state into a circuit that is supported by a particular quantum processor, analyzing it's feasibility with minimal measurement errors itself becomes a critical task. Moreover, obtaining insights on cost of noisy $\rm CNOT$ gates with feasibility certificates for an entangled target gate can also be very useful for quantum processor designer. To this end, QCOpt can be an incredibly useful diagnostic tool for proving \textit{infeasibility} of decomposing any arbitrary unitary using hardware-based elementary gates. Although, proving infeasibility of an MIP can be NP-hard in worst case, there are numerous sophisticated techniques in MIP solvers such as conflict graph analysis, dual proof analysis, bound-tightening presolves and isolation of an Irreducible Infeasible Subset of constraints (IIS) from a large set of constraints \cite{witzig2021computational}. These aforementioned methods can also get very efficient by further reducing the search space of feasibility via valid constraints as described in section \ref{subsec:valid}. Thus, the design philosophy of QCOpt makes it easy to perform this kind of infeasibility analysis, about which we present a result on the classic Toffoli gate in section \ref{subsec:toffoli_infeas}.

\section{Case Studies using QuantumCircuitOpt}
\label{sec:results}
In this section, we provide proof-of-concept case studies to demonstrate the efficacy of using QCOpt for obtaining optimal quantum circuit design. These studies are implemented in QCOpt v0.3.0 using Gurobi v9.1.2 \cite{gurobi} as an underlying MIP solver on an Intel 8 Cores i9 machine running at 2.40 GHz with 32 GB of RAM running a Mac OS. All the input settings for the test target gates in this section are available at this open-source link: \textit{\url{https://github.com/harshangrjn/QuantumCircuitOpt.jl/tree/master/examples}}.

\noindent
\\ \textbf{Validation}: The correctness of the QCOpt's implementation has been thoroughly validated on numerous standard gate decompositions from the literature and IBM's Qiskit library \cite{aleksandrowicz2019qiskit}. Moreover, the open-source nature of the Julia ecosystem makes replicating these results effortless.  

\subsection{Efficacy of QCOpt's mathematical formulations}
\label{subsec:valid_cons_result}
Table \ref{tab:times_valid_cons} summarizes the run times (in seconds) of QCOpt's implementation for decomposing various target gates as listed in the table. Note that the ``Magic basis'' and ``Grover'' gates in this table correspond to the circuits shown in Figures \ref{fig:magic}(b) and \ref{fig:grover}(b), respectively.  The objective function for all these test gates was to minimize the total number of one and two-qubit gates in the resulting decomposition. 

It is clear from Table \ref{tab:times_valid_cons} that the symmetry-breaking valid constraints (see section \ref{subsec:valid}) provide huge speedups in runtimes, with \textit{up to 180 times}, in comparison with the basic MIP formulation derived in section \ref{sec:math_models}. However, note that the optimal solutions obtained by both these methods are identical, because the valid constraints are only speeding up the MIP solver by reducing the feasible search space. We also observed in our experiments that the basic MIP formulation (w/o VCs) stalled at much larger optimality gaps (without convergence) for circuits with $N \geq 3$. Thus, the MIP  models and the valid constraints in combination with the pre-solving techniques discussed in section \ref{subsec:preopt_enhance} make QCOpt an efficient tool for synthesizing up to medium-scale quantum circuits.  

\begin{table}[ht]
    \centering
    \caption{Run time comparisons of QCOpt on decomposing two-qubit circuits to optimality with and without valid constraints (VCs) from section \ref{subsec:valid}. Here, $N_g$ is the total number of gates in the native set and $D$ is the maximum allowable depth.}
    \begin{tabular}{rrrrrr}
        \toprule
        & & & \multicolumn{3}{c}{\textbf{MIP run times} (sec.)} \\ 
        \cmidrule{4-6}
        \textbf{Target gate} & $N_g$ & $D$ & \textbf{w/o VCs} & \textbf{with VCs} & \textbf{Speedup}
         \\ 
        \cmidrule{1-6}
        controlled-Z \cite{aleksandrowicz2019qiskit} & 72 & 4 & 188.4 & 3.9 & 48.3x \\
        controlled-V \cite{aleksandrowicz2019qiskit} & 9 & 7 & 63.8 & 12.2 & 5.2x \\ 
        controlled-H \cite{aleksandrowicz2019qiskit} & 32 & 5 & 31.1 & 4.6 & 6.7x \\ 
        Magic basis \cite{vatan2004optimal} & 72 & 5 & 597.7 & 177.2 & 3.4x \\ 
        iSwap \cite{aleksandrowicz2019qiskit} & 9 & 10 & 170.9 & 6.1 & 28.0x \\
        Grover \cite{abhijith2018quantum} & 21 & 10 & 180.1 & 1.0 & 180.1x \\
        \bottomrule
    \end{tabular}
    \label{tab:times_valid_cons}
\end{table}

\subsection{Compact circuit realizations}
\label{subsec:new_circuit}
We now present new, compact circuit realizations provided by QCOpt on two-qubit circuits for target gates, ``magic basis'' and ``Grover diffusion operator''. The circuits provided here are exact (up to global phase and machine precision) for each of these target gates. Run times for obtaining these realizations are provided in Table \ref{tab:times_valid_cons} (column 5). In this section, ``compressed depth'' of a circuit is the depth obtained by compressing the one-qubit gates on adjacent-depths which are acting on non-identical qubits into a single depth.
\begin{figure}[ht]
    \centering
    \subfigure[(see \cite{vatan2004optimal})]{
	\includegraphics[scale=0.1]{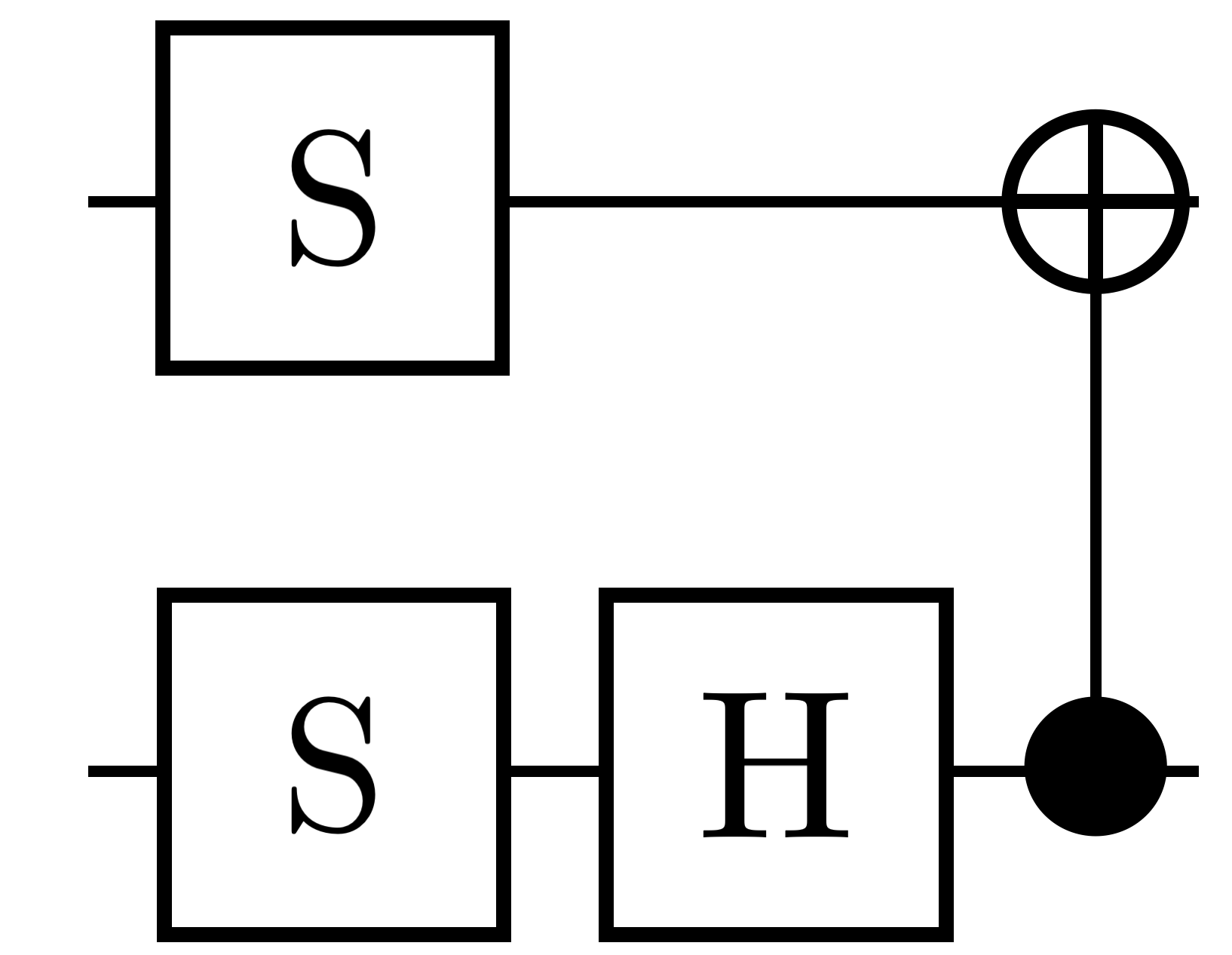}}
	\hspace{1.7cm}
    \subfigure[]{
	\includegraphics[scale=0.095]{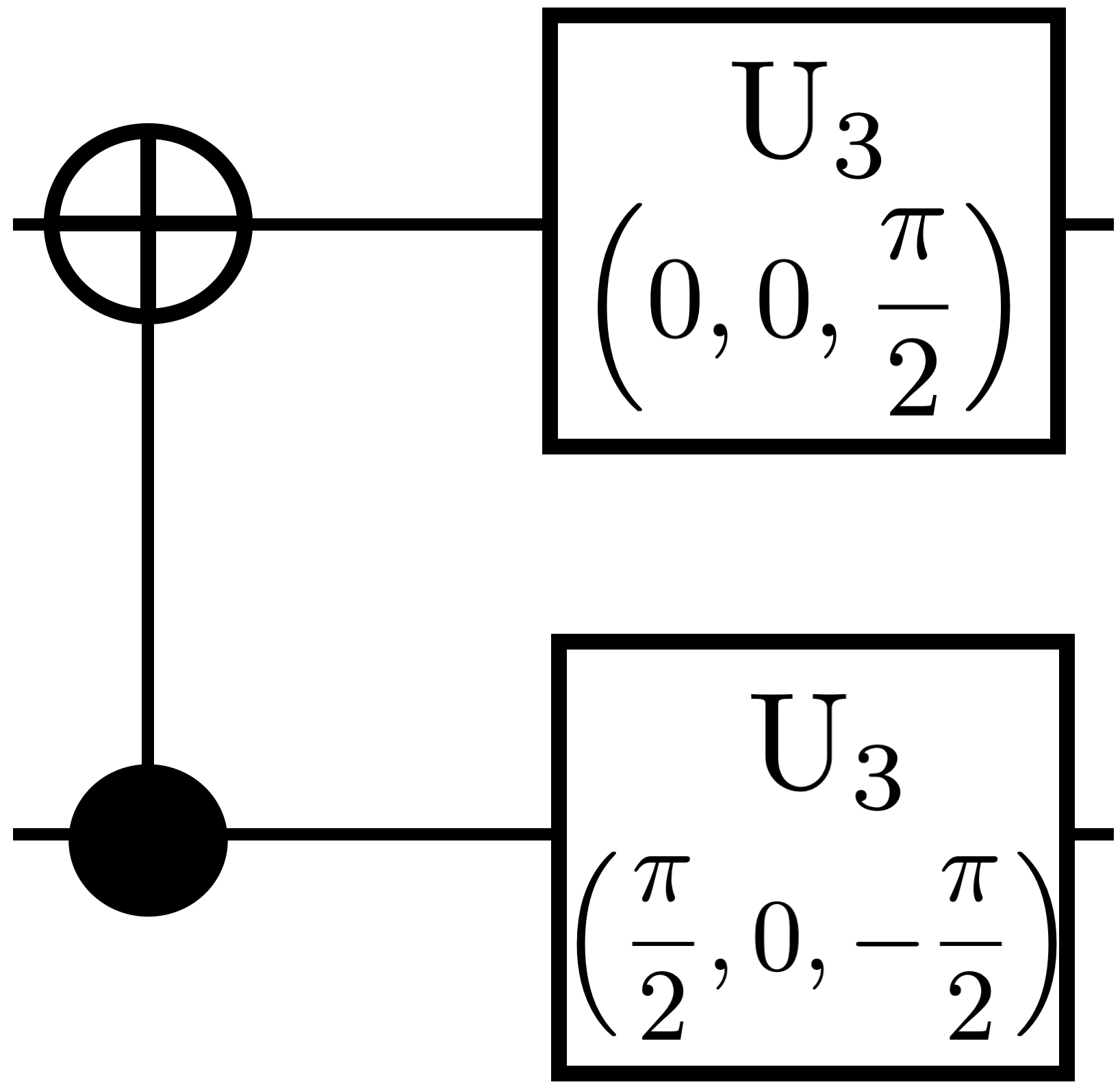}}
	\subfigure[]{
	\includegraphics[scale=0.14]{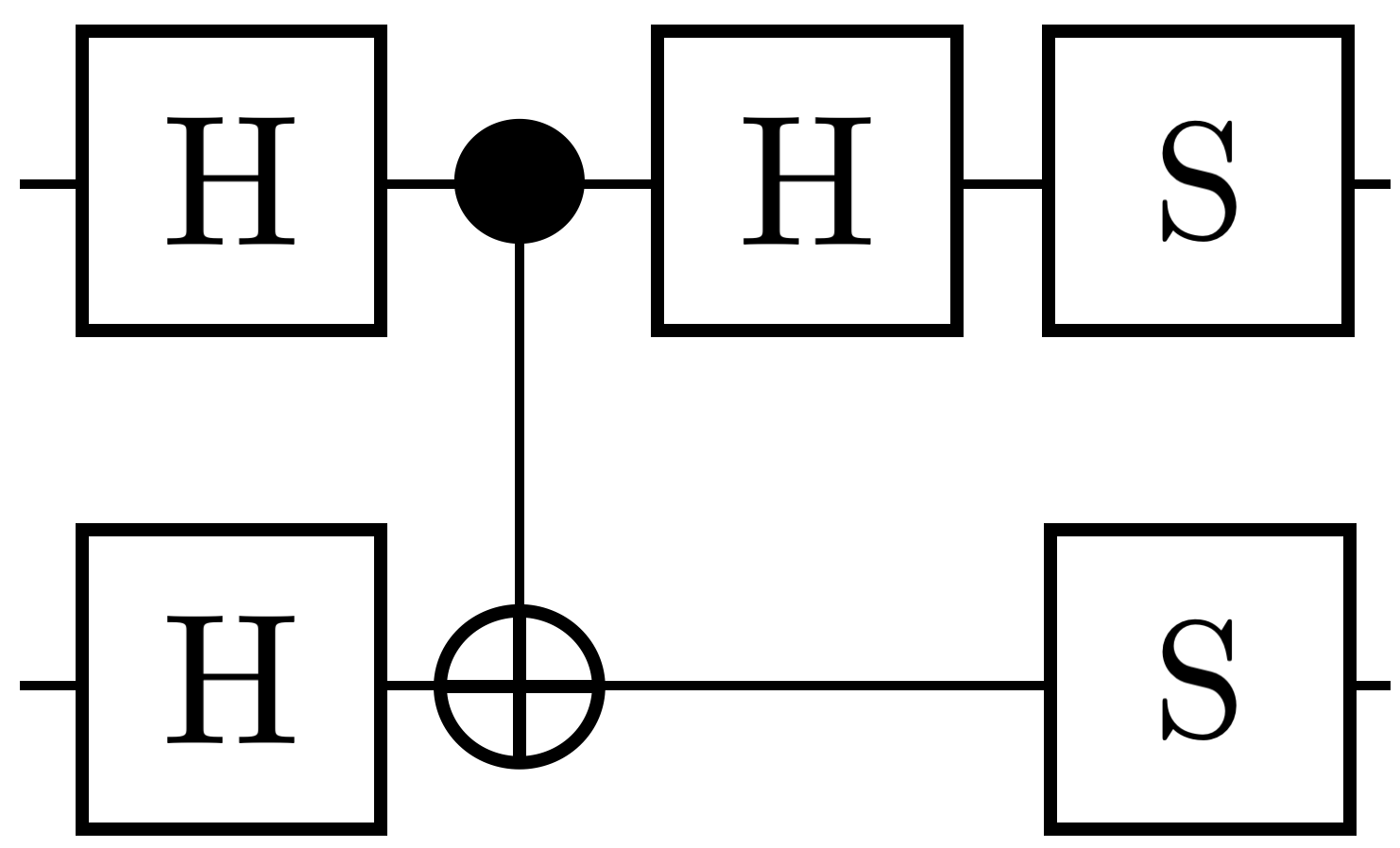}}
	\hspace{0.24cm}
	\subfigure[]{
	\includegraphics[scale=0.09]{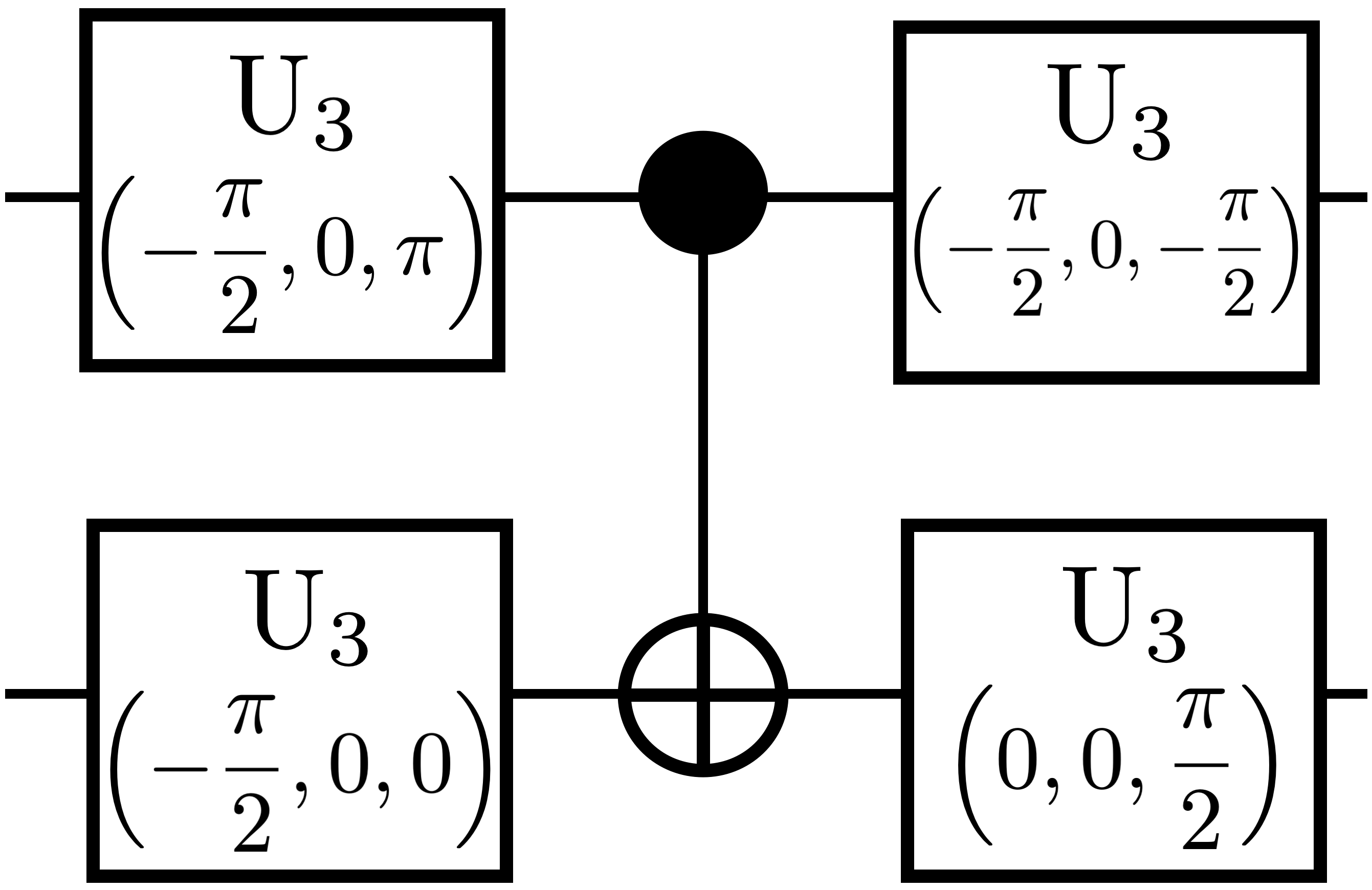}}
    \caption{QCOpt's equivalent circuit realizations for the ``magic basis gate''. (a), (b), (c) and (d) represent optimal circuits using natives gates from the sets $\rm \{S,H,CNOT_{2,1}\}$, $\rm \{U_3, CNOT_{2,1}\}$, $\rm \{S,H,CNOT_{1,2}\}$ and $\rm \{U_3, CNOT_{1,2}\}$, respectively.}
    \label{fig:magic}
\end{figure}

\noindent
\\ \textbf{Magic basis} 
This is an Ising-coupling unitary gate which is natively implemented in trapped-ion quantum computers \cite{hanneke2010realization}. Although there are numerous ways to define the so-called ``magic basis'', we use the one from \cite{hill1997entanglement,vatan2004optimal}. In \cite{vatan2004optimal}, authors provide a way to decompose any two-qubit unitary gate by repeatedly applying the magic basis gate. Theoretical bounds provided in \cite{vatan2004optimal} for implementing any two-qubit unitary gate was based on the magic basis circuit implemented with 4 gates and a depth of 3 as shown in Figure \ref{fig:magic}(a).

Figure \ref{fig:magic} presents four equivalent circuit representations for the magic basis gate with different native gate sets. In the circuit in Figure \ref{fig:magic}(b), obtained using $\rm \{U_3, CNOT_{2,1}\}$ native gates, we observe  $\approx$33.3\% reduction in the compressed depth in comparison with the circuit implemented in \cite{vatan2004optimal} (see Figure \ref{fig:magic}(a)). 

\begin{figure}[ht]
    \centering
	%
	\subfigure[Using $\rm \{H, X, Y, Z, S, T, CNOT_{1,2}\}$ native gates]{
	\includegraphics[scale=0.155]{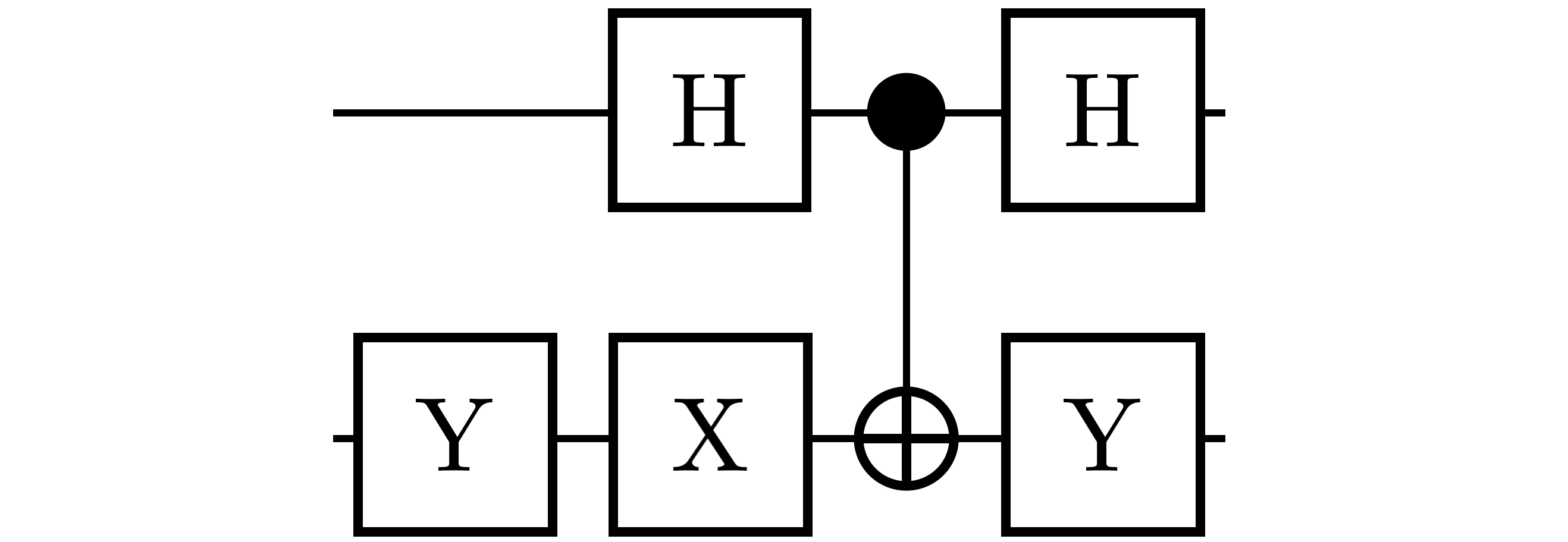}}
	\subfigure[Using $\rm \{U_3, CNOT_{1,2}\}$ native gates]{
	\includegraphics[scale=0.16]{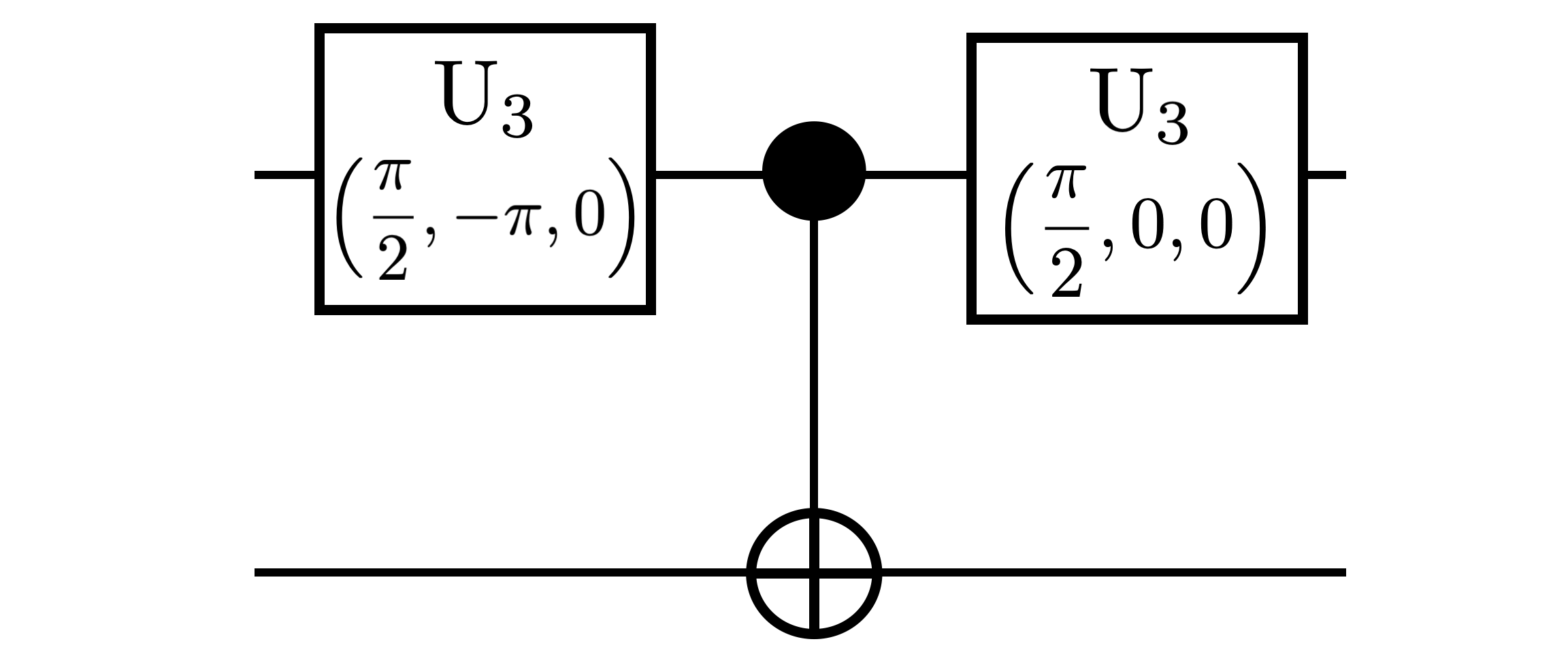}}
    \caption{QCOpt's equivalent circuit realizations for the ``Grover diffusion operator''. (a) and (b) represent optimal circuits using natives gates from sets $\rm \{H, X, Y, Z, S, T, CNOT_{1,2}\}$ and $\rm \{U_3, CNOT_{1,2}\}$, respectively. In particular, (b) proves to be optimal with $\approx$57.1\% reduction in compressed depth in comparison with the circuit provided in \cite{abhijith2018quantum}.}
    \label{fig:grover}
\end{figure}

Implementing an entangling gate like $\rm CNOT_{2,1}$ (like in Figure \ref{fig:magic}(a) and (b)) itself can be expensive, particularly in IBM-type hardware \cite{aleksandrowicz2019qiskit}. It usually necessitates five separate gates using a Hadamard gate on both the qubits and a $\rm CNOT_{1,2}$ gate. Hence, we implemented two equivalent circuits for the magic basis using $\rm \{S,H,CNOT_{1,2}\}$ and $\rm \{U_3, CNOT_{1,2}\}$ native gates, respectively. Clearly, by optimizing using $\rm \{U_3, CNOT_{1,2}\}$ gates, as seen in Figure \ref{fig:magic}(d), QCOpt produces a circuit with one less depth, which also is the most compact way to implement a magic basis gate. To summarize, it is worth to notice that this seemingly simple result in two qubits can provide a substantial improvement in the theoretical bounds for circuit complexity, as derived in \cite{vatan2004optimal}, when the magic basis is repeatedly applied for realizing other two-qubit gates.


\noindent
\\ \textbf{Grover diffusion operator} 
An important part of the Grover's algorithm is the repeated application of the Grover diffusion operator \cite{abhijith2018quantum}. In Figure \ref{fig:grover}, we present two equivalent compact circuit realizations obtained for the Grover operator on two qubits. In Figure \ref{fig:grover}(a), the circuit is optimal with respect to the native gates in the set $\rm \{H, X, Y, Z, S, T, CNOT_{1,2}\}$. In this realization, we observe a reduction of up to $\approx$42.8\% in the circuit's compressed depth, in comparison with the one provided in \cite{abhijith2018quantum}. 

In Figure \ref{fig:grover}(b), the circuit is optimal with respect to the native gates in $\rm \{U_3, CNOT_{1,2}\}$. \textit{To the best of our knowledge, this circuit with three gates is the most compact representation for the Grover operator in the literature}. This realization has $\approx$57.1\% lesser depth in comparison with the circuit provided in \cite{abhijith2018quantum}. Again, to summarize, compact representations for fundamental two-qubit gates can be very beneficial, particularly when a gate like a Grover operator is applied repeatedly for larger problems as discussed in \cite{abhijith2018quantum}.

\subsection{Optimal realizations using  two-qubit gates}
\label{subsec:larger_circuits}
\begin{figure}[ht]
	\centering
	\subfigure[Toffoli gate, quantum cost = 5]{
	\includegraphics[scale=0.14]{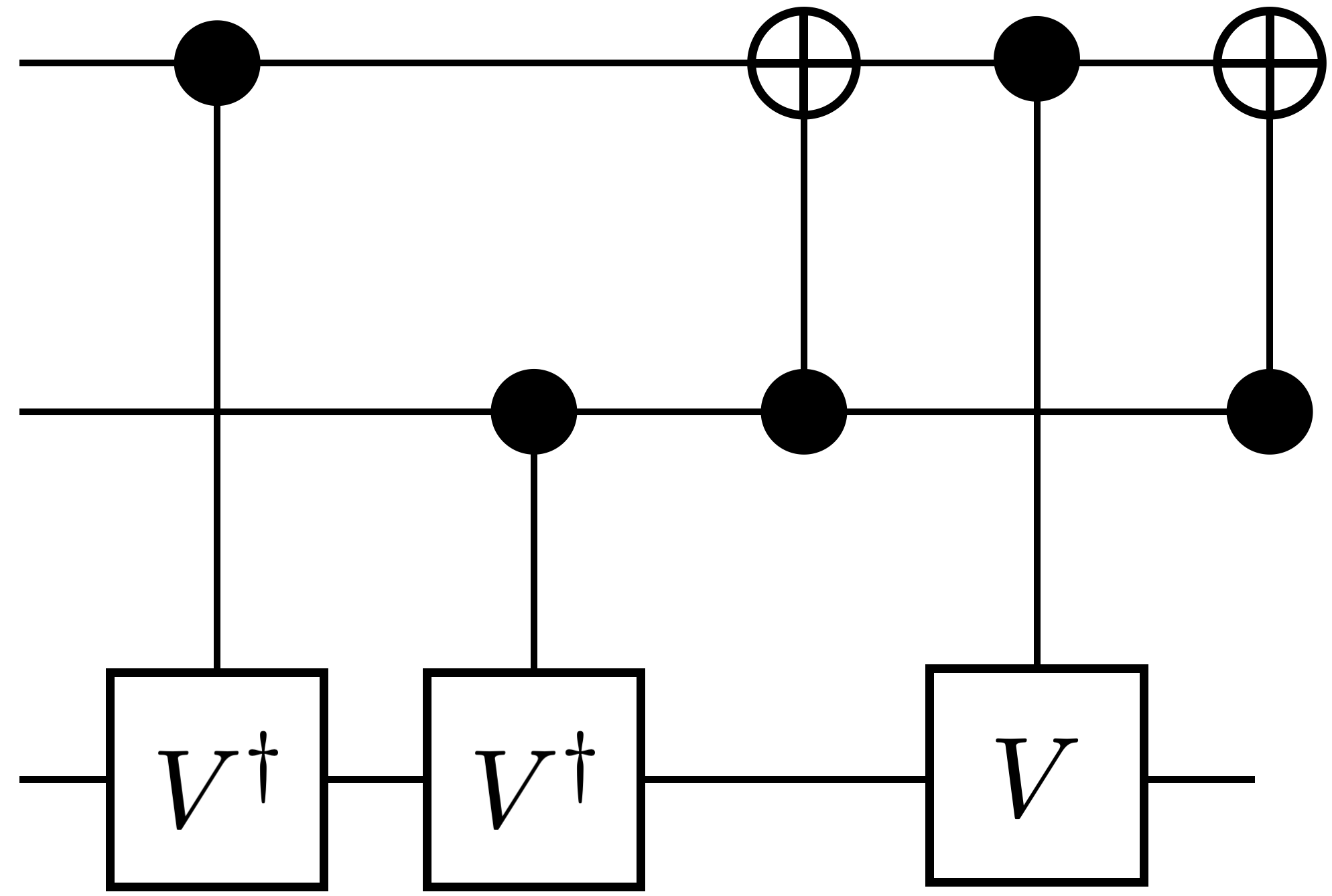}}
	\subfigure[Fredkin gate, quantum cost = 5]{
	\includegraphics[scale=0.138]{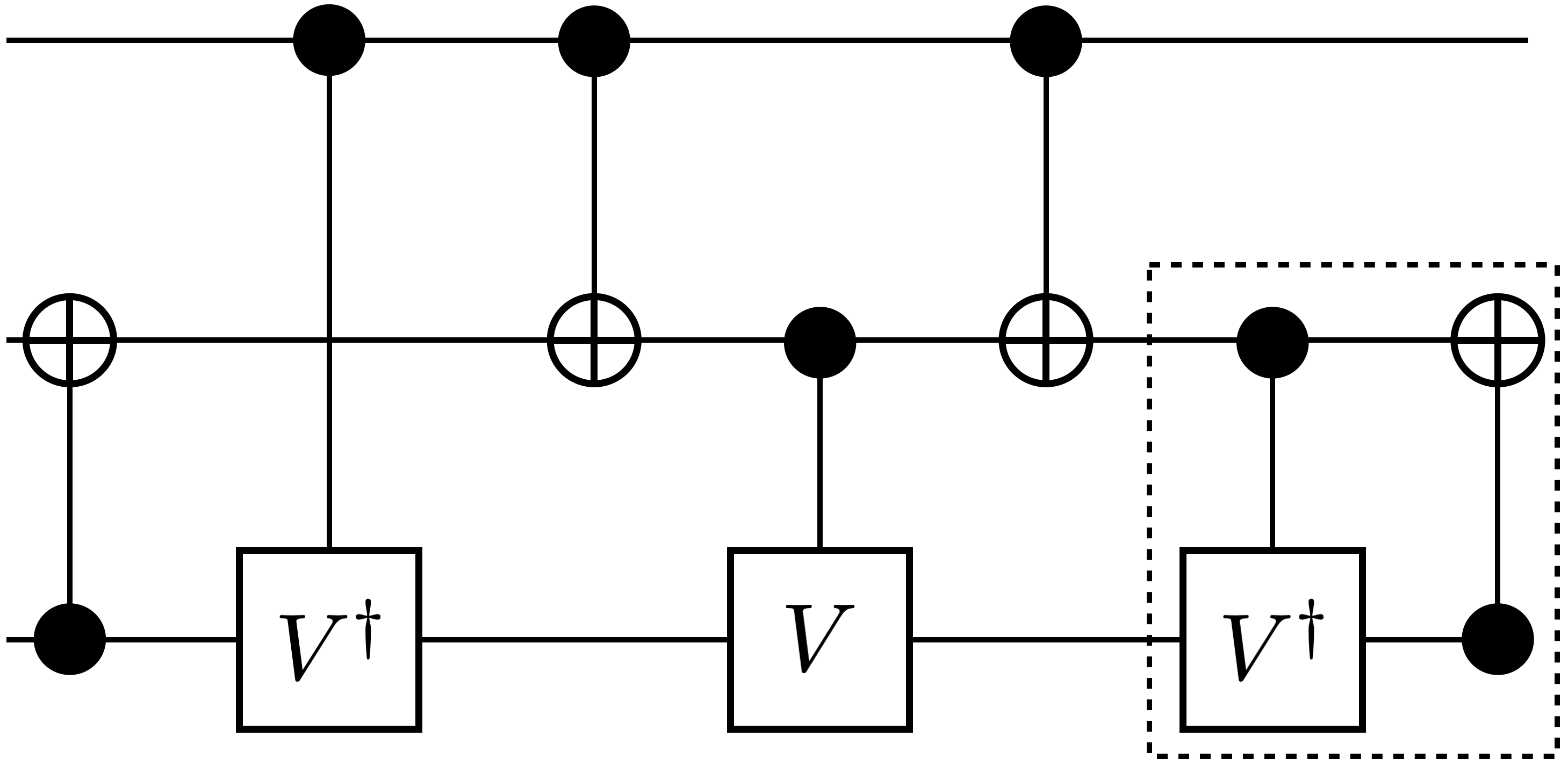}}
	\subfigure[Double-Toffoli gate, quantum cost = 7]{
	\includegraphics[scale=0.14]{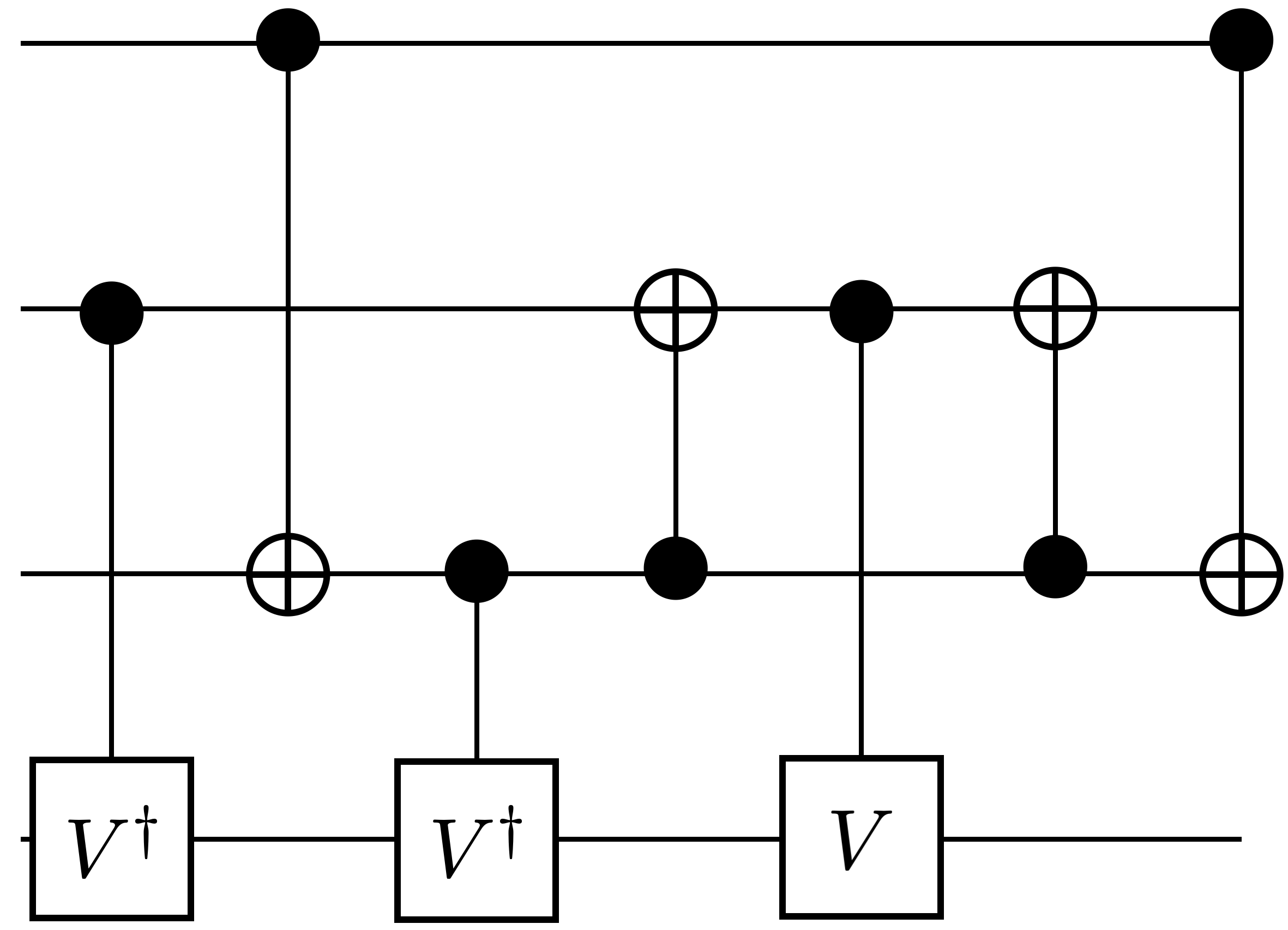}}
	\subfigure[Quantum full-adder gate, quantum cost = 6]{
	\includegraphics[scale=0.16]{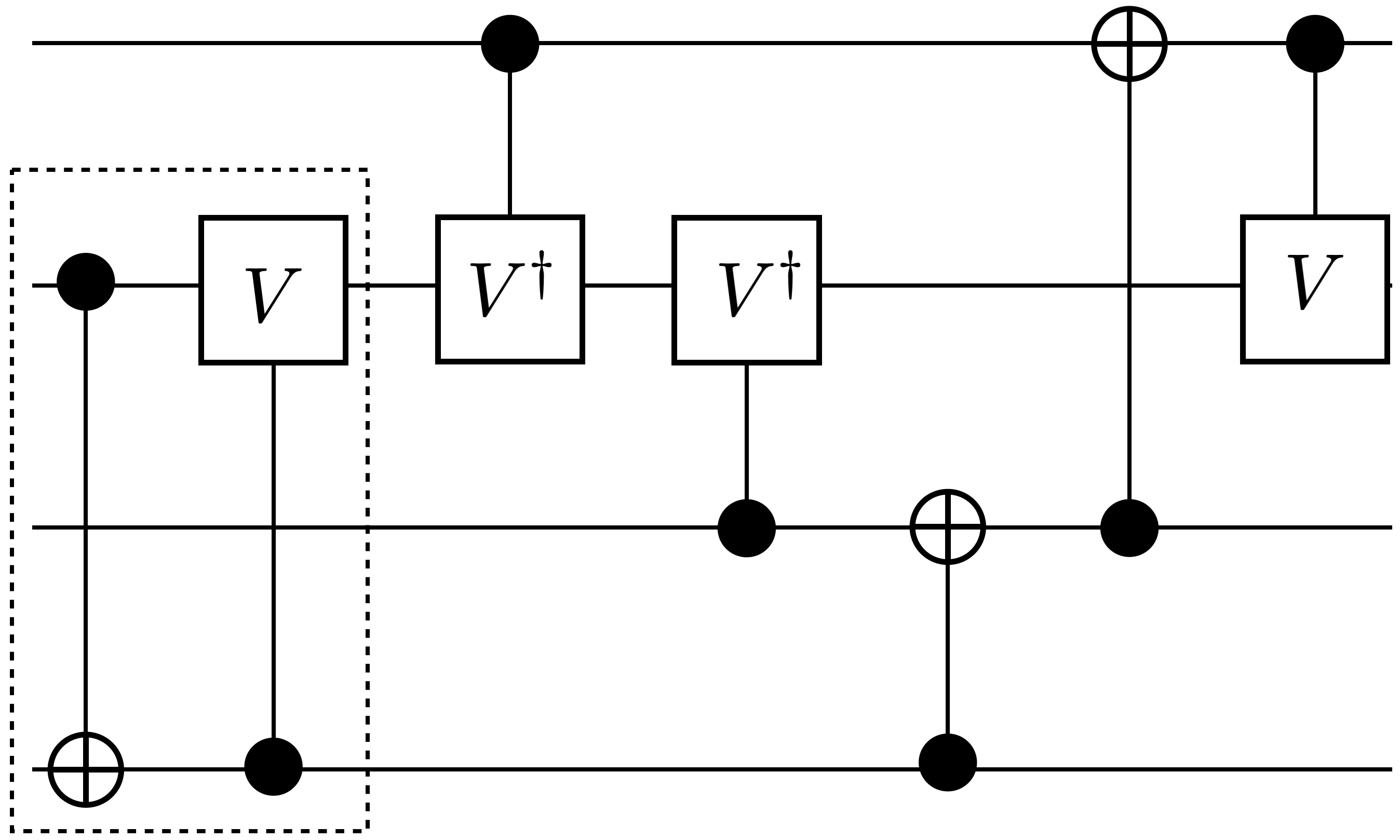}}
	\caption{QCOpt's circuit realizations of three and four qubit circuits. Here, ``quantum cost'' is the cost function described in \cite{smolin1996five}, based on which the adjacent-depth gates on same two qubits are highlighted in dashed boxes.} 
	\label{Fig:larger_circuits}
\end{figure}

Figure \ref{Fig:larger_circuits} represents QCOpt's optimal circuit realizations of target unitarites on three and four qubits, using only two-qubit native gates. We present these gates in comparison with \cite{hung2006optimal}, where the authors implement a brute-force-type enumeration algorithm, based on so-called ``symbolic reachability analysis'', to realize optimal circuits. Thus the circuit realizations in Figure \ref{Fig:larger_circuits} obtained by QCOpt match the ones in \cite{hung2006optimal}. Although, the run times mentioned in \cite{hung2006optimal} were based on older computing platforms, we observe speedups in QCOpt by orders-of-magnitude, due to both improvement in computing platforms and the algorithmic enhancements implemented in QCOpt and state-of-the-art MIP solvers. To be clear, in Figure \ref{Fig:larger_circuits}, QCOpt's run times for decomposing Toffoli, Fredkin, double-Toffoli and quantum full-adder gates are \textit{3.7 s., 72.1 s., 397.4 s., 348.7 s.}, respectively.

It is also noteworthy to mention that authors in \cite{smolin1996five} present an implementation for Fredkin gate and mention the following: ``\textit{However, since the numerical search often gets stuck in local minima, even in cases where it eventually finds a solution, the fact that we were unable to find a smaller implementation of the Fredkin gate is not a proof that one does not exist.}''. Using QCOpt, we prove that the decomposition in \cite{smolin1996five} for the Fredkin gate is indeed globally optimum (in $\approx$75 s.), as shown in Figure \ref{Fig:larger_circuits}(b).

\subsection{QCOpt vs. brute-force enumeration approach}
\label{subsec:brute_force}
The na\"ive approach to decomposition of quantum circuits with optimality guarantees is a brute-force enumeration. This approach tries every possible combination of gates and therefore arrives at the optimal decomposition. To verify every possible circuit realization scales $O({N_g}^D)$ for the native gate set of size $N_g$ and a depth of $D$. Additionally, the evaluation of each circuit  scales with the complexity of matrix multiplication which grows exponentially with the number of qubits in the circuit. Thus, we implemented an approach which goes through all circuit combinations randomly up to a prescribed time limit. For this purpose, we set a time limit of 500 s.

For the Fredkin gate, the above-described approach could not produce any feasible decomposition (fidelity =1) within the time limit for all three randomized trials. However, for the quantum full-adder gate, out of three randomized trials, it produced an optimal decomposition in just one trial (within $\approx$435 s.), and was infeasible for the remaining two trials. For the quantum full-adder gate, as mentioned in section \ref{subsec:larger_circuits}, QCOpt produces an optimal decomposition in less than $\approx$385 seconds. To summarize, in comparison with the above-described na\"ive approach, rigorous mathematical programming-based methods in QCOpt can be incredibly \textit{reliable and robust} for quantum circuit applications.  

\subsection{Proving infeasibility for the Toffoli gate}
\label{subsec:toffoli_infeas}
As discussed in section \ref{subsec:infeas}, QCOpt can not only can prove the existence of an optimal circuit, but can also be used as a tool to prove infeasibility of an existence of an implementation for given target gate. For this purpose, we tested the following: Given a three qubit Toffoli \cite{shende2008cnot} as the target gate, with $\rm \{T_1,T_2,T_3,T^{\dagger}_1,T^{\dagger}_2,T^{\dagger}_3,CNOT_{1,2}, CNOT_{2,3}, CNOT_{1,3}, I\}$ as the elementary gates, and a maximum depth for an allowable decomposition ($D$) equal to $10$, QCOpt provided a proof on infeasibility within $\approx$9 minutes that there does not exist an exact decomposition of depth lesser than or equal to $D$. This again shows the efficacy of mathematical models implemented in QCOpt. 

\section{Conclusions}
\label{sec:conclusions}
In this work, we proposed QuantumCircuitOpt, which implements provably optimal methods, based on mathematical programming models, to compile any arbitrary quantum unitary into a sequence of hardware-native gates. Results indicate that QCOpt can prove the optimality of several medium-sized circuits such as magic basis, Grover diffusion operator, Toffoli, Fredkin, double-Toffoli and quantum full-adder gates, with run times less than a few minutes on a commodity computing hardware.  Moreover, QCOpt produces a new circuit realization for the magic basis gate based on non-trivial angle parameters of the universal gate. We also validated the efficacy of mathematical models in QCOpt and demonstrated it's advantages over a na\"ive brute-force enumeration algorithm.  

To summarize, we believe that the extensibility and modularity of the Julia language and the function-based architecture of the well-documented QuantumCircuitOpt package enable a robust software infrastructure for the continuously evolving research landscape of quantum computing. Furthermore, while we continue enriching the capabilities of QCOpt, we encourage the quantum computing community to explore the package and also contribute to this novel platform.

\section{Acknowledgements}
This work was supported by the U.S. DOE through a quantum computing program sponsored by the Los Alamos National Laboratory (LANL) Information Science \& Technology Institute and the LANL's Laboratory Directed Research and Development (LDRD) program under projects ``20190590ECR: Discrete Optimization Algorithms for Provably Optimal Quantum Circuit Design" and ``20210114ER: Accelerating Combinatorial Optimization with Noisy Analog Hardware''.

\bibliographystyle{IEEEtran}
\bibliography{references.bib}

\end{document}